\documentstyle[12pt]{article}
\renewenvironment{thebibliography}[1]
        {\begin{list}{\arabic{enumi}.}
        {\usecounter{enumi}\setlength{\parsep}{0pt}
         \setlength{\itemsep}{0pt} \settowidth
        {\labelwidth}{#1.}\sloppy}}{\end{list}}

\begin{document}
\vspace*{1cm}
\centerline{\large Sum Rules for $B(M1,0^+_1\rightarrow1^+)$ Strength in
IBM-3 and IBM-4}
\vskip 1cm
\begin{center}
{\bf P.~Halse$^1$, P.~Van Isacker$^1$, and B. R. Barrett$^2$}\\
\bigskip
{\it $^1$GANIL, BP 5027, F-14021 Caen Cedex, France}\\
{\it $^2$Physics Department, University of Arizona, Tucson, AZ 85721, U.S.A.}
\end{center}

\bigskip
\bigskip
\begin{center}
{\small ABSTRACT}
\bigskip

\parbox{13cm}{\small\baselineskip 12pt
Sum rules for $B(M1,0^+_1\rightarrow1^+_i)$ strength are derived for even-even
nuclei in the isospin-invariant forms of the IBM, IBM-3 and IBM-4, in the
cases where the respective natural internal symmetries, isospin $U$(3) and
$U(6)\supset SU$(4), are conserved. Subsequently, the total strength is
resolved into its component partial sums to the allowed isospins (and $SU(4)$
representations in IBM-4). In cases where the usual IBM dynamical symmetries
are also valid, a complete description of all $B(M1,0^+_1\rightarrow1^+_i)$ is
given. In contrast to IBM-2, there is fragmentation of the strength even in
the dynamical symmetry cases, for $T\neq 0$, over two states in IBM-3, and over
three states in IBM-4.  The presence of $pn$ bosons in the ground state of the
extended versions reduces the expected strength from that for IBM-2, allowing
in principle the possibility of using $B(M1)$ data for a given nucleus to
infer which version is the most appropriate.}
\end{center}



\pagebreak
%\bigskip
%\bigskip
\noindent
{\it Introduction}

One of the most important areas of research arising from use of the
proton-neutron Interacting Boson Model (IBM-2) is the investigation of
properties of nuclear levels corresponding to boson states of mixed symmetry
in the $pn$ and orbital degrees of freedom \cite{ibm2}.  (Such structures are
also present in other collective models.)  Prominent amongst these are the
so-called ``scissor'' modes in even-even nuclei, corresponding in the
geometrical models to an oscillation in the angle between the symmetry axes of
the deformed proton and neutron distributions, whose $J^{\pi}$=1$^+$ level is
strongly excited by a largely orbital $M$1 process in ($e,e^\prime$) from the
ground state \cite{expt}.  Recently, an expression for the summed $B$($M$1)
strength in IBM-2 has been derived \cite{ibm2sum}.  It is found to depend upon
the mean number of $d$-bosons in the ground state, and so can be used to
estimate that number from the $B$($M$1) data.
\par
For nuclei where the dominant shell model states involve valence protons and
neutrons in the same orbits, the manifest isospin invariance suggests the
inclusion of this feature also in the IBM.  Two versions have received most
attention: IBM-3 \cite{ibm3}, the minimal isospin invariant model completing
an isospin triplet of $sd$ bosons by the addition of a $T$=1, $M_T$=0
complement (sometimes referred to as $\delta$) to the $\pi$ ($M_T$=1) and
$\nu$ ($M_T$=-1) bosons of IBM-2, and so allowing classification by an isospin
$U$(3) group containing a boson realisation of the usual isospin $SU$(2);
IBM-4 \cite{ibm4}, further augmented by the addition of a $T$=0, $S$=1 boson
(sometimes referred to as $\sigma$), allowing classification by an
isospin-spin $U$(6) group that can be reduced via a Wigner supermultiplet
$SU$(4) to separate $SU$(2) groups for the isospin and spin.
\par
In this letter, we present IBM-3 and IBM-4 $B$($M$1) sum rules for even-even
nuclei in cases where the above internal symmetries are conserved, and
subsequently resolve the total strength into its components to each internal
symmetry representation (isospin, $SU(4)$ label). Finally, for the usual IBM
dynamical symmetries, individual $B(M1)$'s are completely specified. Thus we
obtain various expressions appropriate to the symmetry-limit cases, which may
be used to give ``benchmark'' values as have often proved useful in IBM work.
In addition, these will be seen to allow the possibility of using the
$B(M$1) data to infer which version of the IBM is the most appropriate; the
uncertainties involved are discussed.

\bigskip
\noindent
{\it M1 Sum Rules}

Although the IBM-4 magnetic dipole operator could in principle contain many
terms involving combinations of the various orbital, spin, and isospin
operators, the one-boson analogues of those in the nucleon operator are
expected to be the most important,
\begin{equation}
{\bf T}^{(1)} = \sqrt{\frac{3}{4\pi}}\; (g_{l0} {\bf L} + \frac{1}{2} g_{l1}
\sum_{k} T_0(k) {\bf L}(k) + g_{s0} {\bf S} + g_{s1} {\bf Y}_0) ,
\end{equation}
where ${\bf Y}_{0}$ is the boson analogue of the nucleon operator
$\sum_{k} T_0(k) {\bf S}(k)$, and, for instance,
\begin{equation}
g_{l1} = g_{\pi} - g_{\nu}.
\end{equation}
This restricted form has indeed proved satisfactory in previous
applications \cite{ibm41,ibm42}.

The summed $B(M1)$ strength can be equated to a ground state expectation value
\begin{equation}
\sum_{i} B(M1,0^+_1 \rightarrow 1^+_i) = \langle 0^+_1|
{\bf T}^{(1)}\cdot {\bf T}^{(1)} | 0^+_1 \rangle ,
\end{equation}
where the dot denotes the angular momentum scalar product.
Natural $U(6)\supset SU(4)$ and $T S$ labels for the IBM-4 $N$-boson
isospin-$T$ ground state are [$N$] ($0T0$) $T \; 0$ \cite{ibm4}, in which
case analysis of selection rules reveals that only the (tl).(tl) term in the
product contributes:
\begin{equation}
\sum_{i} B(M1,0^+_1\rightarrow1^+_i) = \frac{3}{16\pi} \; g_{l1}^2
\langle 0^+_1 | \left(\sum_{k} T_0(k) {\bf L}(k)\right)\cdot
\left(\sum_{k} T_0(k){\bf L}(k)\right) | 0^+_1 \rangle .
\end{equation}
The situation for IBM-3, where the natural magnetic dipole operator is
obtained by omitting the spin terms from that for IBM-4, thus differs only in
the labels for the ground state, which are [$N$] $T$ with respect to
$U(3)\supset SU(2)$ \cite{ibm3}.  Indeed, the expression (4) also accommodates
the case of IBM-2, where $\frac{1}{2}T_0$ would be written as $F_0$, and the
ground state carries the (F-spin) $U$(2) label [$N$].  In all cases, we have
\begin{equation}
M_T = N_{\pi} - N_{\nu},\;\; T = |M_T| .
\end{equation}

The assumed total symmetry of the ground state allows the replacement
\cite{ibm41}
\begin{eqnarray}
\left(\sum_{k} T_0(k) {\bf L}(k)\right) \cdot
\left(\sum_{k} T_0(k) {\bf L}(k)\right) & = &
\sum_{k} T_0(k)^2 {\bf L}(k)\cdot {\bf L}(k) + \sum_{k\neq k'} T_0(k) T_0(k')
{\bf L}(k)\cdot {\bf L}(k') \nonumber \\ &
\rightarrow & \frac{1}{N} \sum_{k} T_0(k)^2 \sum_{k} {\bf L}(k)\cdot {\bf L}(k)
\nonumber \\ & + & \frac{1}{N(N-1)} \sum_{k\neq k'} T_0(k) T_0(k')
\sum_{k\neq k'} {\bf L}(k)\cdot {\bf L}(k') .
\end{eqnarray}
Continuing, we have for expectation values in the $L=0, |M_T|=T$ ground state
\begin{equation}
\sum_{k\neq k'} {\bf L}(k)\cdot {\bf L}(k') = {\bf L\cdot L} - \sum_{k} {\bf
L}(k)\cdot {\bf L}(k)
\rightarrow - \sum_l l(l+1) n_l ,
\end{equation}
\begin{equation}
\sum_{k\neq k'} T_0(k)T_0(k') = T_0^2 - \sum_{k} T_0(k)T_0(k)
\rightarrow T^2 - (N - \langle N_{pn} \rangle) ,
\end{equation}
where $N_{pn}$ is the number operator for $pn$ ($M_T$=0) bosons, i.e. $\delta$
and $\sigma$ in IBM-4, $\delta$ in IBM-3, the expectation value being
trivially zero in IBM-2.  Thus we have
\begin{equation}
\sum_{i} B(M1,0^+_1\rightarrow1^+_i) =
\frac{3}{16\pi}\; g_{l1}^2\; \Sigma_l l(l+1) \langle n_l \rangle
\frac{ (N-T)(N+T) - N \langle N_{pn} \rangle} {N(N-1)}.
%\nonumber \\ & = &
%\frac{3}{16\pi}\; g_{l1}^2\; \Sigma_l l(l+1) \langle n_l \rangle
%\frac{4N_{\pi}N_{\nu} - N \langle N_{pn} \rangle} {N(N-1)} .
\end{equation}
In the standard IBM, with only $s$ and $d$ bosons, we then have
\begin{equation}
\sum_{i} B(M1,0^+_1\rightarrow1^+_i)
= \frac{9}{8\pi}\; g_{l1}^2\; \langle n_d \rangle\;
\frac{ (N-T)(N+T) - N \langle N_{pn} \rangle}{N(N-1)} .
\end{equation}
For IBM-2, with $N_{pn}$=0, this reduces to the expression derived previously
by Ginocchio \cite{ibm2sum},
\begin{equation}
\sum_{i} B(M1,0^+_1\rightarrow1^+_i)
= \frac{9}{4\pi}\; g_{l1}^2 \; \langle n_d \rangle
\frac{2N_{\pi}N_{\nu}}{N(N-1)} .
\end{equation}
It is apparent that the inclusion of the $pn$ bosons reduces the expected
$B(M1)$ strength, which opens up the possibility of using data on $B(M1)$'s to
infer which version of the model is the most appropriate; we return to this
point below.

Now consider the derivation of $\langle N_{pn} \rangle$ in IBM-4 and IBM-3;
in fact, it is of interest to calculate the separate values
$\langle N_\delta \rangle$ and $\langle N_\sigma \rangle$ in IBM-4.  It is
convenient to introduce
\begin{equation}
N_\sigma = \frac{1}{2}\; (N - \Delta), \;\; \Delta = N_{(10)} - N_\sigma ,
\end{equation}
\begin{equation}
N_\delta = \frac{1}{3}\; \left(N_{(10)} + \sum_{k} Q_0(k)\right) .
\end{equation}
where $Q_0$ is the isospin quadrupole operator, normalised to have matrix
elements -1, 2, -1 for $\pi,\delta,\nu$ respectively.

The homomorphism $SU(4) \sim SO(6)$ suggests that the structure
$U(6)\supset SU(4)$, as well as $U(3)\supset SO(3)$, should be associated with
a complementary boson quasispin group $SU(1,1)$ \cite{BGWyb}, allowing the use
of reduction formulae in the evaluation of matrix elements.  Indeed, explicit
realisations are given by the canonical forms, where $\Omega$ equals half the
number of internal states, $\Omega$ = 3 and 3/2 for IBM-4 and IBM-3
respectively, and $B^+$ creates the $SO(2\Omega)$ scalar (seniority zero)
pair:
\begin{displaymath}
S_+ = \sqrt{\Omega} \; B^+ = \left\{ \begin{array}{lll}
\sqrt{\frac{3}{2}} \; B^{+(0)} & = \frac{1}{2} \;
{\bf b}^{+(10)}\cdot{\bf b}^{+(10)} & {\rm (IBM-3)} \\
\sqrt{3} \; B^{+(000)(00)} & =  \frac{1}{2} \;
({\bf b}^{+(10)}\cdot{\bf b}^{+(10)} - {\bf b}^{+(01)}\cdot{\bf b}^{+(01)})
& {\rm (IBM-4)} \end{array} \right. ,
\end{displaymath}
\begin{equation}
S_- = (S_+)^+ \; , \; \; \; S_0 = \frac{1}{2} (N+\Omega) .
\end{equation}
Under commutation with the relevant generators, the operators $\Delta$ (in
IBM-4) and $Q_0$ (in IBM-3 and IBM-4) both transform according to the finite
dimensional representations labeled by $(S=1, \; M=0)$, while the states
$| [N] (0T0) \rangle$ and $| [N] T \rangle$ transform according to the
infinite dimensional unitary representations labeled by $(S=(T+\Omega)/2,
M=(N+\Omega)/2)$. Thus, using analytic continuations of the usual
$SU(2)$ 3-$j$ symbols \cite{BGWyb},
\begin{eqnarray}
\langle S M | (1 0) | S M \rangle & = &
(-)^{S-M} \;
\left( \begin{array}{ccc} -S & -S & 1 \\ M & -M & 0\end{array} \right)
/
\left( \begin{array}{ccc} -S & -S & 1 \\ S & -S & 0\end{array} \right)
\times
\langle S S | (1 0) | S S \rangle \nonumber \\ & = &
\frac{M}{S} \; \langle S S | (1 0) | S S \rangle .
\end{eqnarray}
Hence
\begin{eqnarray}
\langle [N](0T0)(T0) | \Delta | [N](0T0)(T0) \rangle & = & \;
\frac{T(N+3)}{T+3} ,\\
\langle [N](0T0)(T0) | Q_0 | [N](0T0)(T0) \rangle & = & \;
- \frac{T(N+3)}{T+3} ,\\
\langle [N]T | Q_0 | [N]T \rangle & = & \;
- \frac{T(N+3/2)}{T+3/2} = - \frac{T(2N+3)}{2T+3} ,
\end{eqnarray}
so that
\begin{equation}
\langle [N](0T0)(T0) | N_\sigma | [N](0T0)(T0) \rangle =
\frac{3(N-T)}{2(T+3)} ,
\end{equation}
\begin{equation}
\langle [N](0T0)(T0) | N_\delta | [N](0T0)(T0) \rangle =
\frac{N-T}{2(T+3)} ,
\end{equation}
\begin{equation}
\langle [N](0T0)(T0) | N_{pn} | [N](0T0)(T0) \rangle =
\frac{2(N-T)}{T+3} ,
\end{equation}
\begin{equation}
\langle [N]T | N_\delta | [N]T \rangle = \frac{N-T}{2T+3} ,
\end{equation}
where Eqn.(22) may also be obtained using the standard matrix elements of
$Q_0$ in the $SU(3)\supset SO(3)$ representations $(N0) \; T$ \cite{BGWyb}.

Thus final expressions for the $B(M1)$ sum rules in IBM-3 and IBM-4 are
\begin{eqnarray}
{\rm IBM-3}: \; \sum_{i} B(M1,0^+_1 \rightarrow 1^+_i) &
= & \frac{9}{8\pi} \; g_{l1}^2 \; \langle n_d \rangle\;
\frac{N-T}{N(N-1)} \left(N+T \; - \; \frac{N}{2T+3}\right) \nonumber \\ &
= & \frac{9}{8\pi} \; g_{l1}^2 \; \langle n_d \rangle\;
\frac{(N-T)(2N(T+1)+T(2T+3))}{(N-1)N(2T+3)}, \\
{\rm IBM-4}: \; \sum_{i} B(M1,0^+_1 \rightarrow 1^+_i) &
= & \frac{9}{8\pi} \; g_{l1}^2 \; \langle n_d \rangle\;
\frac{N-T}{N(N-1)} \left(N+T \; - \; \frac{2N}{T+3}\right) \nonumber \\ &
= & \frac{9}{8\pi} \; g_{l1}^2 \; \langle n_d \rangle\;
\frac{(N-T)(N(T+1)+T(T+3))}{(N-1)N(T+3)}.
\end{eqnarray}

\bigskip
\noindent
{\it Resolution over Isospins and Wigner Supermultiplets}

Comparison of the final state internal symmetry representations contained in
the $U(2\Omega)$ representation $[N-1,1]$ with those arising in the Kronecker
products for the ground state (isospin $T$) and $\sum_k T_0(k){\bf L}(k)$,
yields
\begin{eqnarray}
\rm{IBM-3} : & [N-1,1] & T \; (T\neq 0\; {\rm or}\; N),
                         \; \; \; T+1 \; (T\neq N),\\
\rm{IBM-4} : & [N-1,1] & (0T0) (T0) \; (T\neq 0\; {\rm or}\;N) \nonumber \\
             &         & (1T1) (T0) \; (T\neq 0\; {\rm or}\;N),
                      \; \; \; (T+1,0) (T\neq N).
\end{eqnarray}
The ratios of $B(M1)$ strength to subspaces defined by the representations of
the various symmetry labels, including the experimentally accessible isospin,
involve simply ratios of the group coupling coefficients.

For IBM-3, we require the $U(3)\supset SO(3)$ \cite{u3coeffs} and
Clebsch-Gordan coefficients
\begin{eqnarray}
\frac{\sum_i B(M1,0^+_1\rightarrow(1^+,T)_i)}
     {\sum_i B(M1,0^+_1\rightarrow(1^+,T+1)_i)} & = &
\frac{\langle (N0)T \; (10)1 | (N-1,1)T \rangle^2}
     {\langle (N0)T \; (10)1 | (N-1,1)T+1 \rangle^2} \times
\frac{\langle TT \; 10 | TT \rangle^2}
     {\langle TT \; 10 | T+1 \; T \rangle^2} \nonumber \\ & = &
\frac{T(2T+3)(N+T+1)}{(T+2)N},
\end{eqnarray}
so that
\begin{eqnarray}
\sum_i B(M1,0^+_1\rightarrow(1^+,T)_i) & = &
\frac{9}{8\pi} \; g_{l1}^2 \; \langle n_d \rangle\;
\frac{T(N-T)(N+T+1)}{(T+1)N(N-1)}, \\
\sum_i B(M1,0^+_1\rightarrow(1^+,T+1)_i) & = &
\frac{9}{8\pi} \; g_{l1}^2 \; \langle n_d \rangle\;
\frac{(T+2)(N-T)}{(T+1)(2T+3)(N-1)}.
\end{eqnarray}

\bigskip
For IBM-4, $SU(4)\supset SU(2)\times SU(2)$ and Clebsch-Gordan coefficients
yield
\begin{eqnarray}
\frac{\sum_i B(M1,0^+_1\rightarrow(1^+,(1T1)T)_i)}
     {\sum_i B(M1,0^+_1\rightarrow(1^+,(1T1)T+1)_i)} & = &
\frac{\langle (0T0)T0 \; (101)10 | (1T1)T0 \rangle^2}
     {\langle (0T0)T0 \; (101)10 | (N-1,1)T+1,0 \rangle^2} \nonumber \\
& \times &
\frac{\langle TT \; 10 | TT \rangle^2}
     {\langle TT \; 10 | T+1 \; T \rangle^2} = \frac{3T}{T+4} .
\end{eqnarray}
Ratios involving the $B(M1)$ sum to $(0T0)$ could also be derived, using
in addition coefficients for $U(6)\supset SU(4)$. However, we note that the
complete resolution of the strength can be simply obtained in this case by
evaluating the $B(M1)$ sum to $(1T1)\:T+1$ via the ground state expectation
value of $T^{(1)}\:T_-T_+\: T^{(1)}/2(T+1)$ (for $M_T=+T$); this follows
the derivation presented above. One finds
\begin{eqnarray}
\sum_i B(M1,0^+_1\rightarrow(1^+,(0T0)T)_i) & = &
\frac{9}{16\pi} \; g_{l1}^2 \; \langle n_d \rangle\;
\frac{2T(N-T)(N+T+4)}{(T+4)N(N-1)}, \\
\sum_i B(M1,0^+_1\rightarrow(1^+,(1T1)T)_i) & = &
\frac{9}{16\pi} \; g_{l1}^2 \; \langle n_d \rangle\;
\frac{3T(T+2)(N-T)}{(T+1)(T+3)(T+4)(N-1)}, \\
\sum_i B(M1,0^+_1\rightarrow(1^+,(1T1)T+1)_i) & = &
\frac{9}{16\pi} \; g_{l1}^2 \; \langle n_d \rangle\;
\frac{(T+2)(N-T)}{(T+1)(T+3)(N-1)}.
\end{eqnarray}

\bigskip
\noindent
{\it $B(M1)$'s in Dynamical Symmetries}

In cases where the $sd$ space dynamical symmetries ($U(5), O(6), SU(3)$) are
valid, transitions proceed to at most one orbital representation \cite{ibm2}.
However, from the presentation above it is seen that there is still generally
fragmentation of the M1 strength, in contrast to IBM-2, over two states in
IBM-3, and three in IBM-4, unless $T=0$ when only one transition is allowed in
both versions.  Furthermore, analytic values are available for $\langle n_d
\rangle$ [1],
\begin{eqnarray}
\rm{U(5):} \; \; \langle n_d \rangle & = & 0 ,  \\
\rm{SO(6):} \; \; \langle n_d \rangle & = & {\displaystyle
\frac{N(N-1)}{2(N+1)}} ,  \\
\rm{SU(3):} \; \; \langle n_d \rangle & = & {\displaystyle
\frac{4N(N-1)}{3(2N-1)}} ,
\end{eqnarray}
allowing closed formulae for the various transitions to be obtained by
substitution into the relevant expression (Eqns.(28,29) and (31-33)).

\bigskip
\noindent
{\it Discussion}

Since the above expressions for $M1$ strength differ between IBM-2, -3, and
-4, they furnish a possible means of inferring from $M1$ data which version is
the most appropriate for a given nucleus.  However, it should be noted that:
1) The extended versions are indicated by successively reduced strength, and
so might be wrongly implicated by some $M1$'s being undetected.  2) The
effects of departures from the assumed internal symmetries are not known; in
particular, $SU(4)$ breaking in IBM-4 may lead to involvement of the strong
isovector spin term.  3) Another extension of the IBM(-2), to include
$g$-bosons, leads to increased $M1$ strength \cite{gboson},
which could offset or even
reverse any decrease due to the presence of $pn$ bosons.

\bigskip
\noindent
{\it Acknowledgements}

One of us (BRB) acknowledges the hospitality of GANIL,
where this work was carried out,
and PVI thanks Joe Ginocchio for a discussion concerning this work.
This research is supported in part
by the National Science Foundation, Grant No. INT-9415451
and by the Centre National de Recherche Scientifique.

\pagebreak
%\bigskip
%\bigskip
\noindent
{\bf References}\\[-0.5cm]


\begin{thebibliography}{[10]}
\bibitem{ibm2} F. Iachello and A. Arima,  {\it The Interacting Boson Model},
Cambridge University Press, Cambridge, 1987, and references therein.
\bibitem{expt} A. Richter,  Nucl. Phys. A522 (1991) 139c, and references
therein.
\bibitem{ibm2sum} J. N. Ginocchio, Phys. Lett. B 265 (1991) 6.
\bibitem{ibm3} J. P. Elliott and A. P. White, Phys. Lett 97B (1980) 169.
\bibitem{ibm4} J. P. Elliott and J. A. Evans, Phys. Lett 101B (1981) 216.
\bibitem{ibm41} P. Halse, J. P. Elliott, and J. A. Evans, Nucl. Phys. A417
(1984) 301.
\bibitem{ibm42} P. Halse, Nucl. Phys. A445 (1985) 93.
\bibitem{BGWyb} B. G. Wybourne, {\it Classical Groups for Physicists}, Wiley,
New York, 1974.
\bibitem{u3coeffs} J. D. Vergados,  Nucl. Phys. A111 (1968) 681.
\bibitem{gboson} B. R. Barrett and P. Halse, Phys. Lett 155B (1985) 133.
\end{thebibliography}
\end{document}